\documentclass[useAMS,usenatbib]{mn2e}
\usepackage{graphicx}
\usepackage{txfonts}
\title[Impact of MHD Turbulence on Thermal Wind Balance in the Sun]{Impact of MHD Turbulence on Thermal Wind Balance in the Sun}
\author[Youhei Masada]{Youhei. Masada$^{1,2}$\thanks{E-mail:ymasada@harbor.kobe-u.ac.jp} \\
$^{1}$Department of Computational Science, Kobe University, 657-8501, Kobe, Japan\\
$^{2}$Hinode Science Center, National Astronomical Observatory of Japan, 181-8588, Mitaka, Japan}
\begin{document}
\date{Accepted 2010 November 9. Received 2010 November 2; in original form 2010 October 2}
\pagerange{\pageref{firstpage}--\pageref{lastpage}} \pubyear{2010}
\maketitle
\label{firstpage}
\begin{abstract}
The possible role of magneto-rotational instability (MRI) and its driven MHD turbulence in the solar interior is 
studied on the basis of the linear and nonlinear theories coupling with physical parameters, providing 
solar rotation profile inverted from the helioseismic observation and a standard model for the internal structure 
of the sun. 
We find that the MRI venue is confined to the higher-latitude tachocline and lower-latitude near-surface 
shear layer. It is especially interesting that the MRI-active region around the tachocline closely overlaps 
with the area indicating a steep entropy rise which is required from the thermal wind balance in the sun. 
This suggests that the MRI-driven turbulence plays a crucial role in maintaining the thermal wind 
balance in the sun via the exceptional turbulent heating and equatorward angular momentum transports. 
The warm pole existing around the tachocline might be a natural outcome of the turbulent activities energized 
by the MRI.
\end{abstract}


\section{Introduction}
Two important unsolved issues in solar physics are the mechanisms of magnetic dynamo and angular 
momentum transport in the solar interior. These two processes essentially affect one another 
nonlinearly, which makes the physical processes more difficult to resolve even using huge computer 
facilities. Despite the development of numerical and observational methods, we have not yet drawn a 
complete physical picture of these processes. 

Solar physicists believe that two types of dynamo mechanisms operate in the solar interior. The flux 
transport-type dynamo sustains the global, coherent, magnetic field by the stretching $\Omega$-effect 
around the tachocline (Parker 1993;  Dikpati \& Charbonneu 1999; Dikpati \& Gilman 2006). The convective 
dynamo, in which the convective motion stores free energy, plays a role in generating relatively weak, 
small scale magnetic fields (Parker 1955; Weiss 1994; Miesch et al. 2000; Brun \& Toomre 2002; Brun et al. 
2004). 

The solar rotation profile revealed by helioseismic observation tell us that our knowledge of angular 
momentum transport in the solar interior is still insufficient (Kosovichev 1997; Schou et al. 1998; Thompson 
et al. 2003; Howe 2009). We have no consensus on how the differential rotation is physically developed 
and maintained in the convective zone (e.g., Kitchatinov \& Rudiger 1995; Durney 1999; Rempel 2005; 
Miesch et al. 2006; Balbus 2009; Balbus et al. 2009). 

There is a crucial MHD process related to these outstanding issues, which is  "Magneto-Rotational 
Instability (MRI)" (Balbus \& Hawley 1991, 1994, Menou et al. 2004; Parfrey \& Menou 2007). 
It is essentially of local natures and destabilizes the differentially rotating magnetized system with a 
radially negative shear rate, that is $q \equiv {\rm d}\ln\Omega/{\rm d}\ln\varpi < 0$, where $\Omega$ is 
the angular velocity, and $\varpi $ is the cylindrical radius.  
The typical spatial scale of unstable MRI mode inversely cascades with its linear growth, and it finally operates 
the MHD turbulence. The MRI-driven turbulence is broadly investigated as a 
promising candidate for angular momentum transporter in the astrophysical 
accretion disks (Balbus \& Hawley 1998). 
 
The stability of the solar interior to the MRI is studied at first by Balbus \& Hawley (1994) to understand  
why the solar radiative core rotates uniformly. Menou et al. (2004) construct a linear theory of the MRI in the 
general form which is more suitable for the stellar interior (including heat, magnetic, and viscous diffusivities). 
Recently, Parfrey \& Menou (2007) have pointed out the possibility that the MRI-driven turbulence prevents the 
global coherent dynamo activity in the MRI active higher-latitude tachocline. 

Here we investigate, on the basis of the linear theory, where the MRI should be active in the convective 
zone and tachocline by taking account of the observed rotation profile inverted from the helioseismic data and 
the standard solar structure. Then we discuss a role of the MRI-driven turbulence for maintaining thermal wind 
balance which would be achieved in the solar interior.  
\section{MRI venue in the Solar interior}
We indicate the promising venue for the MRI in the solar interior. Considering axisymmetric WKB plane 
wave perturbation with $\exp[i(k_\varpi \varpi+ k_z z - \omega t)]$ in the Boussinesq approximation, the 
stability of the system to the MRI is governed by the following local dispersion equation, in the cylindrical 
coordinate ($\varpi,\ \phi,\ z$), 
\begin{eqnarray}
&&\frac{k^2}{k_z^2}\tilde{\omega}_{\eta\nu}^4\omega_{\kappa} 
+ \tilde{\omega}_{\eta\nu}^2 \left[\frac{1}{\gamma\rho}(\mathcal{D}\rho)\mathcal{D}\ln P \rho^{-\gamma} \right] 
 \nonumber \\ 
 && +\tilde{\omega}_{\eta}^2\omega_{\kappa} \left[ \frac{1}{\varpi^3} \mathcal{D}(\varpi^4\Omega^2) \right] 
- 4\Omega^2 ({\bf k}\cdot {\bf v}_A)^2\omega_\kappa = 0 \;, \label{eq1}
\end{eqnarray}
where
\begin{eqnarray}
&& \tilde{\omega}^2_{\eta\nu} = \omega_\eta\omega_{\nu} - ({\bf k}\cdot{\bf v}_A)^2 \;, \ \ 
\tilde{\omega}^2_\eta = \omega_\eta^2 - ({\bf k}\cdot{\bf v}_A)^2 \;, \nonumber \\
&&\omega_\chi = \omega + i\chi k^2\ \ (\chi = \kappa , \eta , \nu) \;, \ \ \mathcal{D} \equiv \left( \frac{k_\varpi }{k_z} 
\frac{\partial }{\partial z} - \frac{\partial }{\partial \varpi} \right) \;. \nonumber
\end{eqnarray}
(see Menou et al. 2004 for details). Here $k^2 = k_\varpi^2 + k_z^2$ is the wavenumber, 
${\bf v}_A \equiv {\bf B}/(4\pi\rho)^{1/2}$ is the Alfv\'en velocity,  ${\bf B}$ is the magnetic field, and 
$\gamma $ is the adiabatic index of the gas. The heat, magnetic and viscous diffusivities are represented 
by $\kappa$, $\eta$ and $\nu $ respectively. The others have their usual meanings. 
\begin{figure}
\begin{center}
\scalebox{1.4}{{\includegraphics{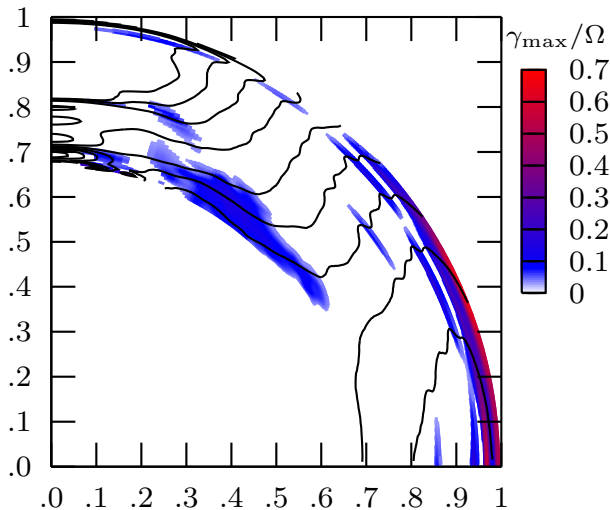}}} 
\caption{The maximum growth rate of the MRI $\gamma_{\rm max}$ which is derived from the dispersion 
equation~(\ref{eq1}) is demonstrated by the color map. Normalization is the local angular velocity $\Omega$. 
The region with the positive growth rate is the promising MRI venue. The rotation profile is overplotted by 
the solid contour line. It increase by $15$ nHz from $330$ nHz to $480$ nHz. Note that the region with zero growth rate 
of the MRI is filled with white. }
\label{fig1}
\end{center}
\end{figure}

Note that the dispersion equation~(\ref{eq1}) is obtained under the condition with stratified hydrostatic 
background, not limited only to the convectively stable background. The convectively unstable mode is also 
included in this dispersion equation. The instability in the form of thermal convection is treated on a footing 
equal to that of rotational instability in this paper as was done by Balbus (2009). The strength of magnetic 
fields in the unperturbed state is assumed to be weak enough that it does not affect the hydrostatic balance. 

We can search the most rapidly growing MRI mode at an arbitrary meridian point ($r$, $\theta$) in the solar 
interior by numerically solving the linear dispersion equation~(\ref{eq1}) coupling with the observed rotation profile 
inverted from the helioseismic data of SOHO Michelson Doppler Imager (MDI) and the standard model for the 
solar internal structure (Christensen-Dalsgaard et al. 1996). When searching for the fastest growing modes, 
we limit their wavelength to being sufficiently smaller than the typical width of the tachocline. 

The promising venue for the MRI is demonstrated by the color map in Figure~1. The color scale represents 
the maximum growth rate of the MRI at each local meridian point. The vertical and horizontal axes denote 
polar and equatorial radii normalized by the solar radius. The rotation profile adopted in our analysis is 
overplotted on the MRI map with solid contour which increases by $15$ nHz from $330$ to $480$ nHz. The 
angular velocity within $\theta \lesssim 10^{\circ}$ is extrapolated by the cubic spline method from the lower 
latitude helioseismic data, where $\theta $ is co-latitude. 

The local structure and the strength of magnetic fields are assumed to be vertical and weak, that is $B_z = 1$ 
G here. The MRI venue is not drastically changed due to the field strength and structure as long as we give
a weak, vertical component of the magnetic field. The destabilizing effect arising from the unstable internal 
gravity wave is eliminated so as to better focus on the destabilizing effect of the MRI modes on the system. 
In contrast, the stabilizing effect due to the density stratification on the MRI is consistently introduced in our analysis.

It is found that, in the solar interior, the MRI-active region is confined to the higher-latitude tachocline and the 
lower-latitude near-surface shear layer. The stabilizing effect due to the density stratification on unstable 
MRI modes is reduced significantly by double diffusive effect in the convectively stable tachocline region 
(Menou et al. 2004; Masada et al. 2007). The MRI venue thus corresponds trivially to the differentially 
rotating zone with negative radial shear $q\equiv  {\rm d}\ln\Omega/{\rm d}\ln\varpi < 0$ here. In addition, 
the wavevector of the most unstable MRI mode tends to be parallel to the gravity vector in the convectively 
stable tachocline as is suggested by Balbus \& Hawley (1994). This is because such modes can grow in 
decoupling with the stabilization effect of the density stratification. When focusing naively on the MRI mode 
with neglecting the effect of the convectively unstable mode, the MHD turbulence powered by the MRI would 
be being operated in these MRI-active regions. 

It would be natural from a physical perspective to discuss the impact of convective motion on the MRI. Since the 
convective mode (unstable mode of internal gravity wave) has a growth rate which is at least an order of magnitude larger 
than that of the MRI,  the profile of large-scale differential rotation drawn using a helioseismic time-averaging method should 
be modified by the vigorous convective motion and not be maintained during the typical evolution time of the MRI. 
The stationary background assumed for deriving the dispersion relation is thus not justified to be applied to the MRI 
in the sun's convective zone. When considering the vigorous convection, the growth of the MRI mode should be 
inhibited in the near-surface shear layer where is one of promising MRI venues. 

In contrast, the tachocline which is a thin transition region between radiative and convection zones generally has 
a convectively stable nature with sub-adiabatic entropy gradient. The large-scale differential rotation should be thus maintained 
without being affected by the turbulent convection during the evolutionary phase of the MRI. 
The dispersion relation adopting the stationary background as unperturbed state is applicable to the MRI mode 
(Parfrey \& Menou 2007). In addition, as is denoted in this section, the stabilization effect due to the sub-adiabatic 
entropy gradient is reduced by diffusive relaxation and does not have a crucial role in suppressing the growth of the MRI. 
We can thus expect that, in the tachocline region, the local scale MRI mode can evolve to the larger scale with leaving its character 
unchanged and finally operates the MHD turbulence with a typical size comparable to the radial width of the tachocline. 
The nonlinear interaction between the convective overshoot and the MRI-driven turbulence is discussed later in \S~3. 

In the following, we focus on the role of the MRI-driven turbulence around the tachocline region. The impact 
of driven MHD turbulence on the thermal wind balance in the solar interior is a special interest of this paper. 
\section{Potential Relation between Thermal Wind Balance and MRI-powered Turbulent Activity}
\subsection{Entropy Rise Resulting from Thermal Wind Balance }
\begin{figure*}
\begin{center}
\begin{tabular}{cc}
\includegraphics[width=0.46\textwidth]{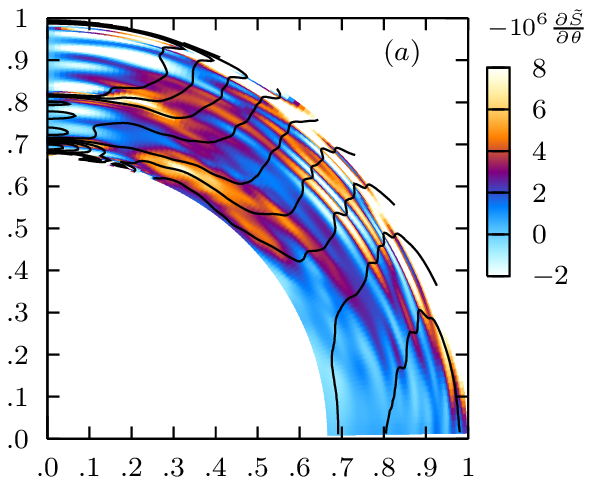} &
\includegraphics[width=0.44\textwidth]{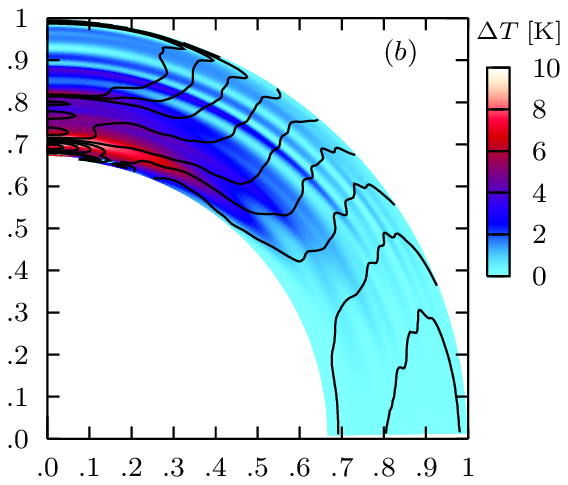}
\end{tabular}
\caption{(a) Specific entropy gradient in $\theta$ direction obtained from the thermal balance 
equation~(\ref{eq2}) coupling with the observed rotation profile. The normalization is chosen as 
$-10^{-6}C_P$ ($\bar{S} = S/C_P$) here. (b) Temperature deviation $\Delta T$ from the equatorial plane
calculated from the thermodynamic relation $\delta T \simeq TC_P^{-1}\int \delta S $. The vertical and horizontal 
axes have same meanings as fig.1. The isorotation contour is overplotted just for the reference on both plots. } 
\label{fig2}
\end{center}
\end{figure*}
The effect of the MRI-driven turbulence on the global coherent dynamo has already been argued by Parfrey 
\& Menou (2007). According to their logic, the coherent stretching process of the magnetic field by the $\Omega$ 
effect is prevented by the MRI-driven turbulence in the higher-latitude tachocline. This implies that the 
flux transport type dynamo efficiently works only in the lower-latitude tachocline. They insist that this prevention 
effect of the MRI-driven turbulence on the dynamo is why the active region is restricted within 
$\sim \pm 45^{\circ}$ around the equator on the solar surface. This is an interesting 
point of view for qualitatively explaining the formation mechanism of the activity belt. 

From a different physical viewpoint, we discuss here the role of the MRI-driven turbulence in 
maintaining the thermal wind balance in the solar interior. The thermal wind equation with a baroclinic term is
\begin{equation}
\varpi \frac{\partial \Omega^2}{\partial z} = \frac{g}{\gamma C_P }\frac{\partial S}{\partial \theta} \;, \label{eq2}
\end{equation}
(Pedlosky 1987; Tassoul 2000) where $S$ is the specific entropy, $g$ is the 
gravity in hydrostatic equilibrium, and $C_P$ is the constant pressure specific heat. This follows from the 
time-steady vorticity equation in conditions with small Rossby numbers.  Equation~(\ref{eq2}) indicates 
that the latitudinal entropy variation drives the differential rotation and yields the deviation from 
Taylor-Proudman state with $\partial\Omega/\partial z = 0$. 

When the observed rotation profile is given, we can directly calculate the left hand side of equation~(\ref{eq2}) 
and derive the entropy gradient in $\theta$ direction. Moreover, we can retread it to the temperature 
deviation by using the thermodynamic relation $\partial T/\partial \theta \simeq \bar{T}C_P^{-1}\partial S/\partial \theta $ 
because of the quasi-incompressible plasma ($\partial P/\partial\theta \simeq 0$), where $\bar{T}$ is the 
reference temperature varying only in the radial direction. This yields information about 
the latitudinal temperature inhomogeneity in the solar interior (e.g., Miesch 2005). 

Figure~2a illustrates the normalized latitudinal entropy gradient $-C_p^{-1}10^{6}\partial S/\partial \theta$ consistent 
with the thermal wind balance. Figure~2b demonstrates the latitudinal temperature deviation 
$\Delta T\ (\equiv \bar{T}C_P^{-1}\Sigma\Delta S\ )$. We choose the equatorial plane as a fiducial location, 
so $\Delta T (\theta =90^{\circ}) = 0$ K in Figure~2b. Note that the temperature obtained from the standard solar 
structure (Christensen-Dalsgaard et al. 1996) is adopted as the reference temperature $\bar{T}$ at each radii. 
The isorotation contour is overplotted on both maps. 

The specific entropy is found to rise steeply toward the pole between $\theta \simeq 20^{\circ}$ and 
$60^{\circ}$ at around the tachocline depth. This suggests that an exceptional heat production is maintained 
in this area. In addition the temperature deviation $\Delta T$ increases with increasing latitude and 
reaches up to $\Delta T \sim 10\ {\rm K} \ $ at the polar region around the tachocline. We interpret  
physically this steep entropic rise and the temperature enhancement around the higher-latitude tachocline 
in relation to turbulent activities powered by the MRI in the following. 

\subsection{MRI-Powered Turbulent Activity}
The higher-latitude tachocline where the entropy steeply rises is reminiscent of the MRI-active region 
denoted in Figure~1 although the governing equations providing each map are completely different. 
Then we attempt to reveal a potential relationship between the steep entropy rise and the MRI-powered 
activities around the solar tachocline. 

A nonlinear outcome of the MRI activity in astrophysical systems is the enhancement of 
turbulent viscous heating. We can expect that the viscous heating maintained by the MRI-driven 
turbulence plays an essential role in inducing the exceptional heat production around the higher 
latitude tachocline which is required from the thermal wind balance. 

In addition the angular momentum tends to be transported equatorward by the MRI-driven turbulence 
in the stably stratified tachocline (Balbus \& Hawley 1994). Then the gaseous matter would advect toward 
the pole with absorbing the heat generated by the MRI-driven turbulence. The warm pole is thus expected 
to be formed at around the tachocline depth as a natural consequence of the MRI-powered turbulent activities. 

For simplicity, we assume that the differential rotation in the tachocline is maintained sufficiently long 
to balance the angular momentum transport entirely in the solar interior. Then we can evaluate 
the turbulent heating rate provided by the MRI-driven turbulence in the tachocline region. 

The Maxwell stress $\mathcal{M}_{\varpi\phi} $ resulted from the MRI-driven turbulence would tap into 
the fraction of the free energy stored in the differential rotation $e_{\rm diff}$ and is given by
\begin{equation}
\mathcal{M}_{\varpi\phi} \equiv - \frac{ B_\varpi B_\phi }{4\pi} = f e_{\rm diff} \;, \label{eq3} 
\end{equation}
where $f$ is an arbitrary parameter ranged from $0.01$ to $0.1$ (Masada et al. in prep). 
The free energy stored in the differential rotation $e_{\rm diff}$ would be written by
\begin{eqnarray}
e_{\rm diff} & = & \rho_0 \int [\Omega^2_{\rm diff} (\varpi)\varpi^2 - \Omega_{\rm rigid}^2 \varpi^2]
{\rm d}{\varpi}{\rm d}\phi {\rm d}z\ \Big{/}\int {\rm d}\varpi {\rm d}\phi {\rm d} z\nonumber \\
                     & \simeq & \rho_0 |q| \Omega_0^2 \varpi_0 \Delta \varpi \;,  \label{eq4}
\end{eqnarray}
where 
\begin{eqnarray}
&& \Omega_{\rm diff}(\varpi)  = \Omega_0\ (\varpi/\varpi_0)^q \;, \nonumber \\
&& \Omega_{\rm rigid}  = \Omega_0\ [(\varpi_0 + \Delta\varpi/2)/\varpi_0]^q \nonumber \;.
\end{eqnarray}
Here $\Omega_{\rm diff}(\varpi)$ is the angular velocity at differentially rotating state with 
maximum free energy and $\Omega_{\rm rigid}$ shows the constant angular velocity at rigidly rotating 
state with minimum free energy. Note that $\varpi_0$ is the fiducial radius, $\Omega_0$ is the angular velocity 
at the fiducial radius,  $\Delta \varpi $ is the radial width of shear layer, and $\rho_0$ is 
the averaged density around $\varpi_0$.  
 
The viscous heating rate resulted from the MRI-driven turbulence is thus, at around the fiducial 
radius $\varpi_0$, 
\begin{equation}
\epsilon  \simeq  \xi \left(\frac{{\rm d}\Omega}{{\rm d} \ln r} \right)^2  =  f \rho_0 |q|^2\Omega_0^3
\varpi_0 \Delta \varpi \;, \label{eq5}
\end{equation}
where $\xi \equiv \mathcal{M}_{\varpi\phi }/|q| \Omega_0$ is the viscous energy deposition rate. Note 
that the turbulent viscosity is then given by $\nu_{\rm turb} \simeq \xi/\rho$. 

Figure~3 shows the viscous heating rate $\epsilon$ which would be sustained by the MRI-driven 
turbulence around the tachocline as a function of co-latitude $\theta$ in the cases $f = 0.01$, 
$0.05$, and $0.1$. The fiducial parameters $\Omega_0$, $q$, $\varpi_0$ and $\Delta \varpi$  are 
systematically determined from the observed rotation profile by averaging over their radial variations. 
Thus, they vary only with the latitude. The averaged density around the tachocline is chosen 
as constant, that is $\rho_0 = 0.2\ {\rm g\ cm^{-3}}$ from the standard solar model. The viscous heating rate 
due to the MRI-driven turbulence is found to reach up to $\mathcal{O}(1)\ {\rm erg\ cm^{-3}\ sec^{-1}}$ with 
plausible physical parameters. 

The typical traveling time $\tau_{\rm travel}$ of the poleward advecting matter due to the turbulent 
angular momentum transport is estimated by the diffusion time,  that is $\tau_{\rm travel} \simeq l^2/\nu_{\rm turb}$, 
where $l$ is the typical size of the MRI-active region measured along the latitude. Then the total energy 
gain $\psi$ of the advecting matter through the MRI-active region can be evaluated as, without depending on 
the arbitrary parameter $f$,
\begin{eqnarray}
\psi & = & \epsilon\tau_{\rm travel} = \rho_0 |q|^2\Omega_0^2 l^2 \nonumber \\
       & \simeq &  5\times 10^8 {\rm erg\ cm^{-3}}\left( \frac{q}{0.5}\right)^2 
       \left( \frac{\Omega_0}{\Omega_{\rm tac}}\right)^2\left( \frac{l}{l_{\rm active}} \right)^2 \;, \label{eq6}
\end{eqnarray}
where $\Omega_{\rm tac} = 2.69\times10^{-6}\ {\rm sec^{-1}}$ is the angular velocity at the tachocline. 
The typical size of the MRI-active region is chosen as $l_{\rm active} = 2\pi \times 0.7 R_{\odot}/8$ here. 
The poleward advection flow is thus heated up by $\Delta T = \psi/(\rho k_B/m) \simeq \mathcal{O}(10) $ K 
during which it goes through the MRI-active region around the tachocline. 

We would like to stress that the typical timescale of the poleward flow traveling the MRI active region becomes
\begin{eqnarray}
&&\tau_{\rm travel}  \simeq 3.2\times 10^8 \ {\rm sec}  \\
&& \left(\frac{f}{0.05}\right)^{-1}\left(\frac{\varpi_0}{0.7R_{\odot}}\right)^{-1}
                              \left(\frac{\Delta \varpi_0}{0.01R_{\odot}}\right)^{-1}\left(\frac{\Omega_0}{\Omega_{\rm tac}}\right)^{-1}
                              \left(\frac{l}{l_{\rm active}}\right)^2 \;. \nonumber
\end{eqnarray}
This is almost comparable to one solar cycle. Then we can predict that the poleward flow resulting from the 
MRI-driven turbulence has a propagating velocity $\mathcal{O}(100)\ {\rm cm\ sec^{-1}}$ 
($\simeq l_{\rm active}/\tau_{\rm travel}$) in the higher-latitude tachocline. 
The physical picture we have arrived at here is more or less speculative, but provides new insight into 
the role of the MHD turbulence driven by the MRI in the solar interior. Further numerical study on nonlinear 
properties of the MRI around the tachocline should be prompted to make this picture robust. 
\begin{figure}
\begin{center}
\scalebox{1.1} {{\includegraphics{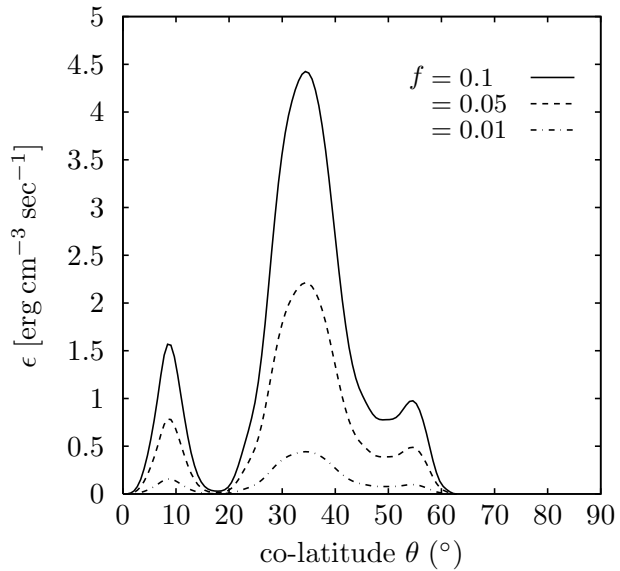}}}
\caption{Turbulent viscous heating rate $\epsilon $ sustained by the MRI-driven turbulence as a function 
of co-latitude $\theta$ in the cases with $f=0.01$, $0.05$, and $0.1$. The region with the turbulent heat 
production closely overlaps the area indicating the entropy rise required from  the thermal wind balance. } 
\label{fig3}
\end{center}
\end{figure}

Based on the current understanding, roughly the upper third of the tachocline is the overshoot layer and the 
remaining lower two-thirds have the same thermal properties as the radiative zone (c.f., Gilman 2000; 
Miesch et al. 2005). It would be thus expected that the MRI-driven turbulence nonlinearly interact with 
the convective overshoot (i.e., overshooting plumes) in the upper third of the tachocline region. 

The impact of the convective hydrodynamic turbulence on the MRI-driven turbulence is studied by 
Workman \& Armitage (2008). They find that the convective turbulence modifies the properties of the MRI-driven 
turbulence with the effect varying depending on the spatial scale of hydrodynamic forcing. Large-scale 
forcing surprisingly boosts the intensity of the MHD turbulence with leaving the character of the MRI 
unchanged. Low amplitude small-scale forcing may modestly suppress the MRI. 

This implies that the nonlinear properties of the MRI in the overshoot layer also depend on the characteristics of the 
overshooting plumes (such as, overshooting velocity, its latitude-dependence and/or filling factor of plumes 
which is the volume occupancy of the turbulent plume in the total volume of the system) although they still 
remain unsettled in the solar interior because the numerical simulation has not yet fully resolved the energy 
dissipation scale (Gilman \& Fox 1997; Brummell et al. 2002; Rempel 2004). The detailed numerical study 
of MRI-driven turbulence impacted by the turbulent overshooting is beyond the scope of this paper, but will 
be a target of our future work.

\section{Summary}
The MRI active venue is clarified on the basis of the linear theory coupling with the observed rotation 
profile and the standard solar model. It is found that the MRI activity is confined to the higher-latitude 
tachocline and the lower-latitude near-surface shear layer. It is especially interesting that the MRI active 
tachocline region closely overlaps with the area indicating the steep entropy rise which is required from the 
thermal wind balance in the solar interior. 

The MRI-driven turbulence would play a crucial role in achieving the sun's thermal wind balance by 
sustaining the exceptional heating and equatorward angular momentum transport. The warm pole with 
$\Delta T \simeq \mathcal{O}(10)$ K would be the natural outcome of the MRI energized turbulent activities
in the tachocline region. 

The impact of the latitudinal entropy inhomogeneity on the solar rotation profile has been investigated  
numerically by adding the mechanical/thermal forcing effect on the bottom of the convective zone (Rempel 2005; 
Miesch et al. 2006). These commonly indicate that the warm pole in the tachocline is key to 
explaining the rotational features of the sun. The MRI energized turbulence could maintain the  
warm pole expected from the thermal wind balance and the numerical studies. 

Recently, Balbus et al. (2009) has found that, by assuming the coincidence between the isorotation surface and 
isocontour of $\delta S \equiv S - S' $,  we can obtain the exact solution of the thermal wind equation~(\ref{eq3}) 
by the characteristic method, where $S$ and $S'$ are regarded as the entropy distribution with and without the 
rotational effects. We stress that the characteristic curves they obtained can correctly reproduce the 
rotation profile of the mid-convective zone (see also Balbus 2009). This implies that, at the quasi-steady state, 
the baroclinic effect has a main role in sustaining the observed rotation profile. 
 
The role of the baroclinic effect is fully controversial at the current status. Although the importance 
of the Reynolds stress in the convective zone is indicated by large scale simulations (e.g., Brun et al. 2004), 
it does not reach consensus yet because the smallest size of the convective cell and the turbulence 
cascading process are inevitably restricted by the grid size even with the current huge computational facilities. 
The global MHD simulation which can capture the localized MRI growth and the other local scale phenomena 
will help us to correctly understand the physical origin of the solar rotation profile and the related 
magnetic dynamo process in the future. 

The rotation profile of the sun reflects much physical information of the solar interior. 
The helioseismic study covering the sun's polar region will reveal further details of the internal 
flow properties and the roles of the MRI in the thermal wind balance of the solar interior. 
\section*{Acknowledgments}
A special note of thanks to T. Sekii who provides Y.M. with a lot of information on the solar interior 
and the inverted MDI data of the solar rotation profile. Y.M thanks to the anonymous referee for 
the constructive and suggestive comments. Y.M. also thank S. Tsuneta, K. Kusano, T. Sano and 
T. Hanawa for scientific discussions and constructive comments.

\label{lastpage}

\end{document}